\documentclass[journal]{IEEEtran}

\usepackage[utf8]{inputenc}
\usepackage{color}
\usepackage{array}
\usepackage{verbatim}
\usepackage{float}
\usepackage{amsmath}
\usepackage{amsthm}
\usepackage{amssymb}
\usepackage{graphicx}
\usepackage{longtable}

\usepackage{verbatim}
\usepackage{float}
\usepackage{epstopdf}
\usepackage{subfigure}
\usepackage{enumerate}
\usepackage{lmodern}
\usepackage{graphicx}
\usepackage{amsmath,mathrsfs,amssymb}
\usepackage{caption}
\usepackage{array}
\usepackage{longtable}
\usepackage{algpseudocode}
\usepackage{amsmath}
\usepackage{float}
\usepackage{algorithm}
\usepackage{booktabs}
\usepackage{array}

\usepackage[unicode=true,
 bookmarks=false,
 breaklinks=false,pdfborder={0 0 1},colorlinks=false]
 {hyperref}
\hypersetup{
 colorlinks,bookmarksopen,bookmarksnumbered,citecolor=blue,urlcolor=blue}
\usepackage[]{apacite}
\makeatletter
 \let\oldforeign@language\foreign@language
 \DeclareRobustCommand{\foreign@language}[1]{%
   \lowercase{\oldforeign@language{#1}}}

 \let\oldforeign@language\foreign@language
 \DeclareRobustCommand{\foreign@language}[1]{%
   \lowercase{\oldforeign@language{#1}}}

\ifCLASSINFOpdf
\else
\fi

\begin{document}
%
%
\onecolumn
\noindent\rule{18.1cm}{2pt}\\
\underline{To cite this article (APA Style):}
{\bf{\textcolor{red}{Hashim, H. A., Ayinde, B. O., \& Abido, M. A. (2016). Optimal placement of relay nodes in wireless sensor network using artificial bee colony algorithm. Journal of Network and Computer Applications, 64, 239-248.}}}\\
\noindent\rule{18.1cm}{2pt}\\

{ \bf The published version (DOI) can be found at: \href{http://dx.doi.org/10.1016/j.jnca.2015.09.013}{10.1016/j.jnca.2015.09.013} }\\

\vspace{40pt}Please note that where the full-text provided is the Author Accepted Manuscript or Post-Print version this may differ from the final Published version. { \bf To cite this publication, please use the final published version.}\\

\textbf{
	\begin{center}
		Personal use of this material is permitted. Permission from the author(s) and/or copyright holder(s), must be obtained for all other uses, in any current or future media, including reprinting or republishing this material for advertising or promotional purposes.\vspace{60pt}\\
	\end{center}
	\begin{flushleft}
		\underline{Publication information:}\\
 Date of Submission: December 2014.\\
Date of Acceptance: September 2015.\\
Available Online: February 2016.\vspace{280pt}
\end{flushleft} }
\scriptsize{ \bf
	\vspace{20pt}Please contact us and provide details if you believe this document breaches copyrights. We will remove access to the work immediately and investigate your claim.
} 

\normalsize

\twocolumn
\title{Optimal Placement of Relay Nodes in Wireless Sensor Network Using Artificial Bee Colony Algorithm}

\author{Hashim A. Hashim$^*$, Babajide. O.~Ayinde, and~Mohamed~A.~Abido
\thanks{$^*$Corresponding author, H. A. Hashim is with the Department of Systems Engineering, King Fahd University of Petroleum and Minerals, Dhahran, 31261, Saudi Arabia, e-mail: hmoham33@uwo.ca}
\thanks{B. O. Ayinde is with the Department of Systems Engineering, King Fahd University of Petroleum and Minerals, Dhahran, 31261, Saudi Arabia.}
\thanks{M. A. Abido is with the Department of Electrical Engineering, King Fahd University of Petroleum and Minerals, Dhahran, 31261, Saudi Arabia.}
}


\markboth{--,~Vol.~-, No.~-, \today}{Hashim \MakeLowercase{\textit{et al.}}: Optimal Placement of Relay Nodes in Wireless Sensor Network Using Artificial Bee Colony Algorithm}
\markboth{}{Hashim \MakeLowercase{\textit{et al.}}: Optimal Placement of Relay Nodes in Wireless Sensor Network Using Artificial Bee Colony Algorithm}

\maketitle

\begin{abstract}
Deploying sensor nodes randomly most of the time generates initial communication hole even in highly dense networks. These communication holes cannot be totally eliminated even when the deployment is done in a structured manner. In either case, the resulting inter-node distances may degrade the performance of the network. This paper proposes an enhanced deployment algorithm based on Artificial Bee Colony (ABC). The ABC-based deployment is guaranteed to extend the lifetime by optimizing the network parameters and constraining the total number of deployed relays. Simulations validate the effectiveness of the proposed strategy under different cases of problem complexity. Results show that the proposed approach improves the network lifetime considerably when compared to solutions reported in the literature such as Shortest Path 3-D grid Deployment (SP3D) algorithm.
\end{abstract}


\IEEEpeerreviewmaketitle{}

\section{Introduction}

%
%
%
%
\IEEEPARstart{W}{ireless} sensor networks (WSNs) have tiny, low-powered and multi-functional nodes. Through collaborative efforts, these nodes have the capabilities of performing complex tasks. Simultaneously maximizing both the connectivity and lifetime of WSN from the design standpoint is considered a challenging task \cite{Turjman:2012, Chang:2012}.
Relay nodes (RNs) are generally employed to reduce the energy utilization of the sensor nodes. Owing to the fact that relay nodes are fabricated using powerful transceivers, they have better transmission and reception capabilities \cite{Song:2010, Cheng:2008}. It is worthy of note that, the cost of powerful transceivers and complex embedded circuitry makes RNs much more expensive than normal sensor nodes.\\*[.3pc]
\indent
In some applications, relay nodes are not only forwarding and collecting data in the horizontal plane but also in the y-z plane \cite{Turjman:2012,Al-Turjman:2010}. So, maximizing the lifetime of a WSN with constraints on both cost and connectivity has been shown to be a challenging task especially in 3-D deployment - where nodes are arranged in three-dimensional pattern \cite{Son:2006,Song:2010}. The overall deployment cost is minimized if less number of RNs is deployed. In a large 3-D space, previous work shows that the deployment of nodes is not a trivial task because the search space is wide. Each position yields different connectivity levels and hence, the optimization technique has to be selected carefully to ensure convergence \cite{Bari:2007, Yang:2013}. Most of the time, heuristic algorithms are introduced to improve the solution and the computational efficiency \cite{Turjman:2012}.\\*[.3pc]
\indent
Artificial intelligent(AI) approaches, especially those that are biologically inspired, are commonly used nowadays to solve many complex engineering problems. In the last few decades, different engineering problems have been solved successfully using evolutionary techniques. More recently, a new metaheuristic optimization approach which was first inspired in 2005 and was later modified into what is known today as Artificial Bee Colony (ABC) \cite{Karaboga:2005,Karaboga:2007}.\\*[.3pc]
\indent
The major contribution of this work is to enhance the network lifetime by proposing an ABC-based two-phase relay node deployment strategy in 3-D space called Improved Lifetime Deployment subject to Cost Constraint (ILDCC). The placement problem addressed in this work has been shown in \cite{Bari:2007} to be NP-hard, and locating approximate solution is NP-hard as well \cite{Efrat:2008}. This complexity can be circumvented using a two-phase, two-layered approach. In the first phase, the backbone of the network is connected using minimum number of relay nodes for cost efficiency \cite{Turjman:2012}. In the second phase, a novel approach using heuristic method for searching the global optima is introduced. The parameters of the network are optimized in such a way that the minimum objective function is guaranteed and the desired network connectivity is maintained. The proposed technique shows its effectiveness in solving the placement problems associated with WSNs, and the solution can be adopted in wide range applications such as: WSN for Volcanic Monitoring; relay node deployment in forests to detect fires and report wild life activities; $CO_2$ flux monitoring and imagery; many other outdoor monitoring applications where sensor networks may work under harsh environmental conditions.
\\*[.3pc]
\indent
The rest of this paper is organized as follows: Section \ref{sec2} discusses the state of the art literature review. Section \ref{sec3} presents the problem formulation . Section \ref{sec4} gives a detailed presentation of the proposed deployment strategy and how it is mapped with ABC. Section \ref{sec5} discusses the implementation of the proposed strategy and the simulation results. Section \ref{sec6} concludes the paper and highlights some future works.
\section{Related Work} \label{sec2}
Recently, the effectiveness of deploying relay nodes in wireless sensor networks has been widely investigated. In \cite{Younis:2008}, classification of deployments into random and grid-based are highlighted by the authors. The grid-based deployment yields more accurate positioning and data measurements because nodes are arranged on the grid vertices. Whereas, in random deployment, nodes are randomly scattered and are organized in an ad hoc manner. In 3-D settings, Grid deployment is used because it simplifies the placement problem. In \cite{Cerpa:2004}, connectivity problems were addressed using node redundancy. Redundant nodes are switched off and isolated from the network when idle. However, some of the redundant nodes are turned on to repair connectivity whenever there is network partition. The use of nodes mobility was proposed in \cite{Marta:2009} to repair the disconnected network.\\*[.3pc]
\indent
Likewise in \cite{Lee:2006,Lee:2010}, an optimized approach for connecting disjointed WSN segments was proposed and achieved by modeling the deployment area as a grid with equal-sized cells and carefully populating few relay nodes. The network is then optimized such that, fewest count of cells are selected to be populated by relay nodes till all segments are connected. This is considered a non-deterministic polynomial-time hard (NP-hard) problem. The two layer procedure was basically used to reduce the complexity of the deployment strategy from an NP-hard to a more realistic one \cite{Turjman:2012, Abbasi:2009}. It must be remarked that, the major task of the sensor nodes is to collect and transmit data to the closest Cluster Head (CH) or RN. Hence, the sensor nodes occupy the first layer of the architecture. In this way, the energy consumption of sensor nodes will be considerably minimized as the nodes go back to sleep immediately after transmitting data.\\*[.3pc]
\indent
In the first phase of deployment, the Minimum Spanning Tree (MST) is first used to construct the network backbone by placing minimum number of first phase relay nodes (FPRNs) on 3-D grid vertices. These FPRNs set up a connection among the pre-allocated cluster heads and the base station. In the second phase, extra relay nodes are randomly and densely deployed very close to the backbone devices until the desired connectivity is achieved.\\*[.3pc]
\indent
The Optimized 3-D deployment with Lifetime Constraints (O3DwLC), an improved version of SP3D, was later proposed in  \cite{Turjman:2012} to enhance the lifetime of the network. O3DwLC also implements MST to initialize the connection of the backbone. However, a Semi-Positive Definite optimization (SPD) algorithm was proposed in the second phase of the deployment. Even though constraints were put on cost and lifetime, the algorithm efficiently enhances the overall connectivity of the network. While the SP3D and O3DwLC approaches aim at improving the connectivity of the network, the proposed approach in this work focuses on maximizing the useful lifetime of the wireless sensor network.\\*[.3pc]
\indent
Artificial intelligent approaches have also been used to optimize the lifetime of the network. Multi-objective territorial predator scent marking algorithm was presented to maximize both the coverage and the connectivity of the network using the least energy consumption \cite{Abidin:2014}. ABC and Particle Swarm Optimization algorithms are used to solve placement problem and maximize the lifetime \cite{Mini:2014}. In addition, ABC based algorithm was proposed to determine the optimal 3-D position that satisfies $k$-coverage and $Q$-coverage criteria and subsequently, to extend the lifetime and the coverage of the WSN \cite{Udgata:2011,Ajayan:2013}. To recover partitioned WSNs, a distributed relay node positioning approach was proposed in \cite{Senturka:2014} using virtual force-based movements of relays and Game Theory. Also in \cite{Liu:2014}, an ant colony optimization with greedy migration mechanism was proposed to deploy sensor nodes in order to  maximize coverage and minimize the deployment cost.\\*[.3pc]
\indent
Unlike \cite{Udgata:2011,Mini:2014,Abidin:2014,Liu:2014} which focus on finding the optimal placement of sensor nodes, the proposed solution in this study focuses on extending the network lifetime with optimal deployment of relay nodes. The proposed approach makes the deployment algorithm less complicated when compared with existing solutions with similar objectives.\\*[.3pc]
\section{Problem Formulation} \label{sec3}
\subsection{Assumptions}
A two-layer hierarchical structure is considered in this paper to address the heterogeneous nature of wireless sensor network. The network mainly comprises of devices such as  sensor nodes, cluster heads, relay nodes and the base station. The sensor nodes are localized in the lower layer where their primary role is to sense the targeted phenomena in form of data. The measured data is then sent to RN or CH in the upper layer of the architecture. In this work, the design approach is adopted such that sensor nodes minimize their energy usage by transmitting sensed data over short distance and go to sleep mode when not sensing or transmitting.  The upper layer of the architecture consists of cluster heads, relay nodes and base station. The upper layer devices are equipped with better capabilities and can transmit and receive data over longer distances \cite{Turjman:2012}. They also convey the data from lower layer devices periodically to the base station. \\*[.3pc]
\indent
In order to drift the attention away from the lower layer, it is assumed in this work that sensor nodes have enough resources to perform their tasks. This allows us to pay more attention to the upper layer devices. It was established in \cite{Santamaria:2010, Olariu:2006} that, the cost of sensing the environment and encapsulating it into a packet is much smaller than that used in transmitting and receiving such packets. Thus, the energy dissipated during the communication phase is only taken into consideration. Besides, the two-layer architecture facilitates the decoupling of the energy consumptions in the upper and the lower layer. Other assumptions made in this paper are:
\begin{itemize}
	\item All nodes are static
	\item A periodic data gathering application where data is sensed and is sensed and transmitted by each sensor to its cluster head (CH) and from the CH to another CH or relay node
	\item Multi-hop communication
	\item All CHs have the same transmission range
	\item All relay nodes have the same transmission range of $"r"$ units and
	\item Ideal Media Access Control (MAC) layer with no collisions and retransmissions.\\*[-1pc]
\end{itemize}
\subsection{Energy Consumption}
The general energy consumption model proposed in \cite{Xu:2010, Rodrigues:2007} is adopted in this paper. The energy consumed by the receiver ($J_{rx}$) is given by:
\begin{equation} \label{MyEq1}
\begin{split}
J_{rx} &= L\beta
\end{split}
\end{equation}
Energy consumed by transmitter ($J_{tx}$),
\begin{equation} \label{MyEq2}
\begin{split}
J_{tx} &= L(\epsilon_1+\epsilon_2 d^{\gamma})
\end{split}
\end{equation}
where $\beta$, $\epsilon_1$ and $\epsilon_2$ are designed hardware parameters of a particular transceiver \cite{Turjman:2012}, and L is the packet length, $d$ is the Euclidean distance between the transmitter and receiver, $\gamma$ is the path loss exponent.
Remaining energy $E_{r}$, is given by:
\begin{equation} \label{MyEq3}
\begin{split}
E_{r} &= E_i-TJ_{tx}-RJ_{rx}-AJ_a
\end{split}
\end{equation}
where T, R and A are the arrival rates of the transmitted, received and aggregated packets per round respectively, $E_{i}$ is the initial energy of each node. $J_{tx}$, $J_{rx}$ and $J_a$ are the energy per unit time consumed by each node for a single packet transmission, reception and aggregation respectively.\\*[.3pc]
\indent
The energy model adopted in this paper incorporates the effect of the traffic handled at the relay nodes by virtue of their relative positions to the sink. It must be remarked that the relay nodes near the sink deplete their energy more quickly. This is because they relay data to the sink on behalf of other nodes more often than the others. This eventually results in network partitioning and consequently degrades the network performance. Therefore, the total energy a node consumes is thus given by:
\begin{equation} \label{MyEq4}
\begin{split}
E_{p} &= TJ_{tx}+ kRJ_{rx} + AJ_{a}
\end{split}
\end{equation}
where $k=$ traffic handled by an RN due to data from other  nodes.\\*[-1pc]
\subsection{Network Lifetime Definition}
Lifetime of a wireless sensor network is the time from deployment of relay nodes to the instant when the network is partitioned. Network partitioning in WSN occurs when the algebraic connectivity of the network falls below a threshold solely because some RNs/CHs have drained their energy and can no longer relay data. From the two-layer architecture proposed in this paper, it valid to assume that the entire network will be disconnected by disconnecting most of the relay nodes. In other words, the connectivity of the backbone is centered around the activity of the relay nodes. Using this lifetime definition, the value of the energy's cut-off threshold may theoretically vary from network to network due to factors such as density of nodes and location of nodes within a network. However, the lifetime definition adopted in this work excludes such factors, but centers around the cost of running the backbone elements (RNs/CHs).  In setting the cut-off threshold, the total energy in the backbone of the network is considered, not the energy in the individual nodes. This method eliminates the problem of threshold variation within the network. However, the chosen threshold can vary from application to application, and also depends on the capabilities of the electronics (for example, the transceiver rating) used in the backbone elements.
\subsection{Problem Statement}
Given the Euclidean locations of sensor nodes, cluster heads and base station on a grid wireless sensor network; the aim is to determine the optimal locations of relay nodes that maximize the lifetime of the network while the constraints on cost budget and connectivity are satisfied. In a nutshell, given a sensing task with pre-specified locations of sensor nodes, cluster heads and base station, the proposed solution gives the grid positions of the second phase relay nodes (SPRN) that enhance the overall network lifetime are determined lifetime.
\section{The Proposed Deployment Strategy} \label{sec4}
By considering all the aforementioned challenges associated with 3-D deployment, an effective approach of deploying relay nodes using an adaptive, heuristic non-deterministic algorithm called the ABC is studied in this work. By using a two two-layer architecture called ILDCC-Improved Lifetime Deployment subject to Cost Constraint, the NP-Hard settings can be circumvented. In this work, a cubic grid model proposed in SP3D is adopted. However, the arrangement of the vertices can assume many regular shapes ranging from cubical to pyramidal
\subsection{First Phase of the ILDCC}
\begin{figure}[H]
	\centering
	\includegraphics[scale=0.5]{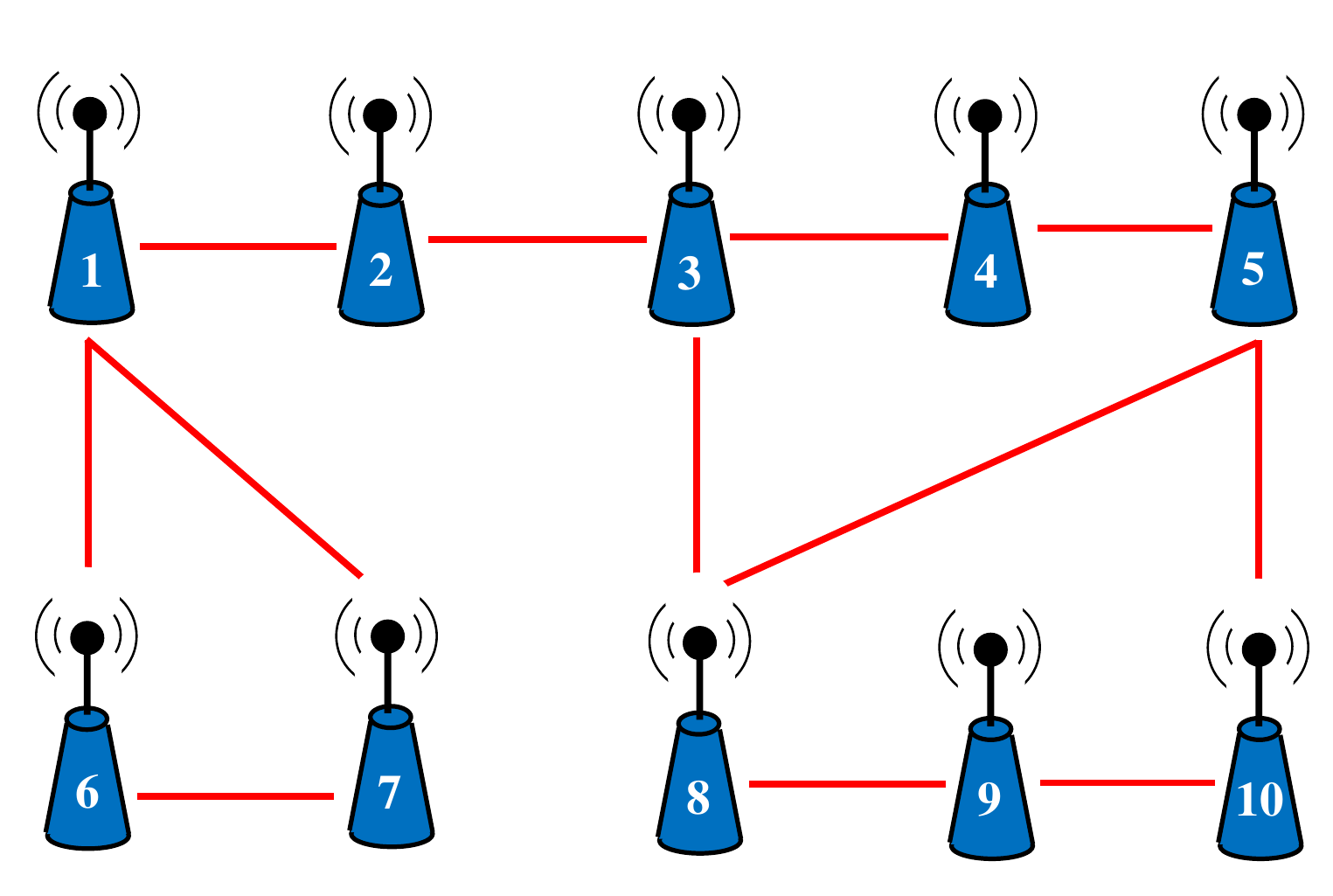}
	\caption{A graph of 10 nodes and 12 links}\label{laplacian}
\end{figure}

For the purpose of qualitatively comparing the proposed algorithm with SP3D, the backbone of the network is also constructed with MST using the First Phase Relay Nodes (FPRN). The SP3D strategy is used due to its efficient ability to maintain a predefined lifetime and select the minimum number of relay nodes needed to construct the network backbone. In addition, SPD3 is well known for harsh environmental applications such as in forest fire detection and soil experiments. The MST ensures that all the second layer devices (relay nodes, cluster heads and the base station) are connected to establish communications.\cite{Turjman:2012}. The MST is implemented using the Algorithm~\ref{alg:MST} as executed in \cite{Turjman:2012}. \\*[.3pc]
\begin{algorithm}
	\caption{MST to construct the connected back-bone $B$}
	\label{alg:MST}
	\begin{algorithmic}[1]
		\Function{Construct B}  {IS: Initial Set of nodes to construct $B$}
		\State {\bf Input} :
		\State {A set IS of the CHs and BS nodes.coordinates};
		\State {\bf Output} :
		\State A set CC of the CHs, minimum RNs, and BS coordinates forming the network Backbone;
		\State {\bf begin}
		\State {CC $=$ set of closest two nodes in IS;}
		\State CC $=$ CC $\cup$ minimum RNs needed to connect them on the 3-D grid;
		\State IS $=$ IS - CC;
		\State CC = CC [ minimum RNs needed to connect them on the 3-D grid;
		\State $N_d$ $=$ number of remaining IS nodes which are not in CC;
		\State $i = 0$;
		\State {\bf for} each remaining node $n_i$ in IS
		\State Calculate $M_i$: Coordinates of minimum number of RNs required to connect $n_i$
		with the closest node in CC.
		\State $i = i + 1$;
		\State {\bf end}
		\State $M = {Mi}$
		\State {\bf while} $Nd > 0$ do
		\State SM $=$ Smallest $M_i$;
		\State CC$=$ CC $\cup$ SM $\cup$ $n_i$;
		\State IS $=$ IS-$n_i$;
		\State M $=$ M - $M_i$;
		\State $N_d $=$ N_d - 1$;
		\State {\bf end}
		\EndFunction
	\end{algorithmic}
\end{algorithm}
\\*[.2pc]
\indent
The backbone $B$ can be perceived as a connected graph ($G$) from the knowledge of graph theory and a semi-definite matrix called the Laplacian matrix ($L$) can be generated \cite{Ghosh:2010,Bhardwaj:2001,Boyd:2006}. $L$ is a symmetrical two-dimensional array that has $-1$ at the element $(i,j)$, if nodes i and j are connected. It has a positive integer number at the element (i,i) denotes the number of edges connected to the node \cite{Turjman:2012}. Eq.~(\ref{MyEq5}) shows the ordering of the eigenvalues of $L$
\begin{equation} \label{MyEq5}
\begin{split}
0=\lambda_1 \leq \lambda_2 \leq \lambda_3 \leq ....\lambda_{n-1} \leq \lambda_n\\
\end{split}
\end{equation}
It can be remarked that, $\lambda_2 = 0$ if and only if Backbone ($G$) is totally disconnected. In effect, $\lambda_2$ is designated as the algebraic connectivity (or the Fielder value) of $G$. The spanning tree algorithm can be used as a tool to tune the network connectivity level to a nominal value by increasing the number of FPRNs. On the other hand, $\lambda_2 = 1$ signifies connection between all nodes in the network. Figure~\ref{laplacian} illustrates an example of a connected network backbone with a size $N=10$. The backbone can be seen as a graph with 10 nodes and 12 links. It must be noted that the removal of a single node can partition the whole network and eventually degrade the overall performance. A typical example of this cornerstone node is the node 3 in Figure~\ref{laplacian}. Eq.(\ref{MyEq5a}) gives the Laplacian matrix ($L$) generated from Figure~\ref{laplacian} with $\lambda_2=0.1764$. The second smallest eigenvalue ($\lambda_2$) of the matrix indicates the number of link count required to partition the network \cite{Turjman:2012}.

\begin{equation} \label{MyEq5a}
L_{10}= {\left(\begin{smallmatrix} 3 & -1 & 0 &0&0&-1&-1&0&0&0\\ -1 & 2 & -1 &0&0&0&0&0&0&0\\ 0 & -1 & 3 &-1&0&0&0&-1&0&0\\0 & 0 & -1 &2&-1&0&0&0&0&0\\0 & 0 & 0 &-1&3&0&0&-1&0&-1\\-1 & 0 & 0 &0&0&2&-1&0&0&0\\-1 & 0 & 0 &0&0&-1&2&0&0&0\\0 & 0 & -1 &0&-1&0&0&3&-1&0\\0 & 0 & 0 &0&0&0&0&-1&2&-1\\0 & 0 & 0 &0&-1&0&0&0&-1&2 \end{smallmatrix}\right)}
\end{equation}
The following definitions in \cite{Merris:1990} is adopted in this work:
\begin{enumerate}\label{Enume1}
	\item Let $G$ be a graph with vertex set $V(G)$ and edge set $E(G)$. The distance $d_G(u,v)$  between two vertices $u,v \in V (G)$ is the minimum number of edges on a path in $G$ between $u$ and $v$.
	\item The distance $d_G(v)$ of a vertex $V$ is the sum of all distances between $v$ and all other vertices of $G$, the Wiener index $W(G)$ of $G$ is given by:
\end{enumerate}
\begin{equation} \label{MyEq6}
\begin{split}
W(G) &= \tfrac{1}{2}\sum_{v\in V(G)}d_G(v)=\tfrac{1}{2}\sum_{(u,v)\in V(G)}d_G(u,v)\\
\end{split}
\end{equation}
Where $d_G(v)$ is the sum of distances between node $v$ and all other vertices of $G$ and $\tfrac{1}{2}$ compensates for the fact that each path between $u$ and $v$ is counted.
$\mu(G)$ is the average distance  between the vertices of $G$ and it is given by:
\begin{equation} \label{MyEq7}
\begin{split}
\mu(G) &= \tfrac{W(G)}{\tfrac{n}{2}}=\tfrac{W(G)}{\left(
	\begin{array}{c}
	\lvert V(G)\rvert\ \\
	2 \\
	\end{array}
	\right)
}\\
\end{split}
\end{equation}
Eq.(\ref{MyEq7}) above is re-written as given in Eq.(\ref{MyEq7}) to compensate for the deviations in the average distance from the real value. Therefore, the proposed solution is robust and can fit into many practical scenarios.
\begin{equation} \label{MyEq7b}
\begin{split}
\mu_{w} (G) &= \tfrac{W(G)}{\tfrac{n}{2}}=\tfrac{W(G)}{\left(
	\begin{array}{c}
	\lvert V(G)\rvert\ \\
	2 \\
	\end{array}
	\right)
}+\Delta\mu(G)\\
\end{split}
\end{equation}
$\Delta\mu(G)$ is selected carefully enough to approximate a practical scenario. Occurrences such as collisions and retransmission can be adapted easily into our solution by lumping all the energy wasted during these activities into the uncertain parameter ($\Delta\mu(G)$). However, the upper bound of this uncertainty must be established in order to guarantee a good solution. Therefore, our approach can be adopted in real world applications. Also, let $T$ be a tree of n-vertices and let the eigenvalues of the corresponding Laplacian matrix be ordered as $\lambda_1 \leq \lambda_2 \leq \lambda_3 \leq ....\lambda_{n-1} \leq \lambda_n$. Then the Wiener index can be computed as
\begin{equation} \label{MyEq8}
\begin{split}
W(G) &= n \sum_{2}^n\tfrac{1}{\lambda_i}\\
\end{split}
\end{equation}
\subsection{Second Phase of the ILDCC}
In this phase, the position of the SPRNs is optimized with the objective of minimizing $\mu_w(G)$ of the backbone by putting a constraint on the number of SPRNs that can be deployed. The cost has a direct relationship with the total number of SPRNs deployed. The fewer the number of SPRNs deployed, the better the deployment solution.
\begin{equation} \label{MyEq9}
\begin{split}
\min \mu_{w}(L(\alpha)) \\
s.t \ \ \ \ \sum_{i=1}^{n_c} \alpha_i \ \ \ where \ \ \ \ \alpha_i\in (0,1)
\end{split}
\end{equation}
\begin{equation} \label{MyEq10}
\begin{split}
\lambda_{2,FPRN+SPRN} \geq \lambda_{2,FPRN}
\end{split}
\end{equation}
\begin{equation} \label{MyEq11}
\begin{split}
L(\alpha)=L_i+ \sum_{i=1}^{n_c}\alpha_i A_i A_i^T
\end{split}
\end{equation}
where $A_i$ $=$ incidence matrix that results from the addition of SPRNs, $L_i$ $=$ initial Laplacian matrix generated using MST algorithm, $n_c$ $=$ candidate positions the SPRN can be placed and $\alpha_1$ is either 1 or 0 depending on whether RN is placed on vertex $i$ or not accordingly.
\\*[.4pc]
Using ILDCC approach, the total number of rounds the network can stay operational is derived. Let the initial rounds ($I_R$) be the total number of rounds after the deployment of the first phase relay nodes, the following mathematical derivations can be carried out:\\
\begin{equation} \label{MyEq12}
\begin{split}
I_R= \tfrac{B_1}{\sum_{p=1}^{FPRN}E_p}
\end{split}
\end{equation}
where $B_1$ $=$ total energy in the network using FPRN
\begin{equation} \label{MyEq13}
\begin{split}
Lifetime= T_R= \tfrac{\sum_{i=1}^{2}B_i}{\sum_{p=1}^{FPRN+SPRN}E_p}
\end{split}
\end{equation}
where
\begin{equation} \label{MyEq14}
\begin{split}
E_{p} = TL(\epsilon_1+\epsilon_2 \mu_{w} ^{\gamma})+ kRL \beta + AJ_{a}
\end{split}
\end{equation}
and $B_2$ =  extra energy due to the addition of SPRN. Therefore, the overall lifetime of the network (total number of rounds) is thus formulated as:
\begin{equation} \label{MyEq15}
\begin{split}
Lifetime= T_R= I_R + \sum_{i=1}^{n_c}\alpha_i E_R
\end{split}
\end{equation}
where $E_R =$ extra rounds due to the addition of the SPRN

\subsection{Implementation of the ABC Algorithm}
\begin{figure}[h]
	\centering
	\includegraphics[scale=0.6]{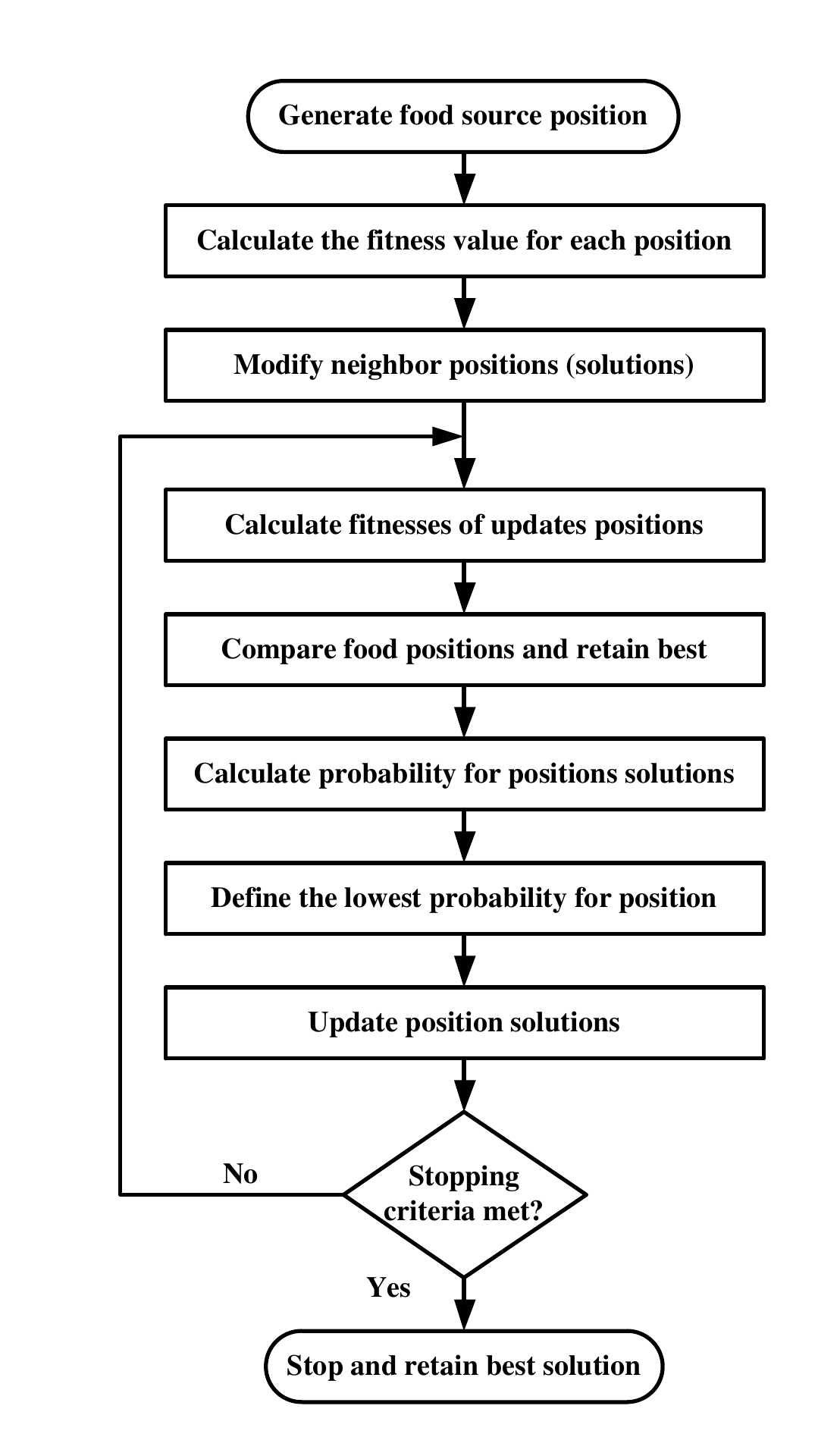}
	\caption{Artificial Bee Colony Algorithm.}\label{abc}
\end{figure}

ABC was initially inspired by the work proposed in \cite{Karaboga:2005} as a new meta-heuristic optimization approach. The ABC algorithm got its inspiration from the way life is structured in the colony of the natural bees. The bees in the colony are usually divided into three groups: employed, onlooker and scout bees. The employed bees search randomly for food and they tag the best position of food as the optimal solution. In order to relay information about the food source and the amount of nectar to other bees, the employed bees dance. The job of the onlookers is to differentiate between the good source and the bad source using dance length, dance type and speed of shaking. They also use these information to determine the quality of food. The scout bees are chosen from the onlooker bees prior a new search for food. Depending on the quality of food, onlooker and scout bees may decide to be employed or vice versa. The similarities between bees food searching and ABC is extensively discussed in \cite{Karaboga:2005,Karaboga:2007}.\\*[.3pc]
\indent
In the ABC algorithm, employed and onlooker bees are responsible for searching in space for the optimal solution while scout bees control the search process as detailed in \cite{Karaboga:2007}. The solution of the optimization problem is the position of the food source while the amount of nectar with respect to the quality is termed as the objective function of the solution. The ABC can be summarized in the flowchart shown in Figure \ref{abc}.
The position of the food source in the search space can be described as follows:
\begin{equation} \label{MyEq15a}
\begin{split}
x^{new}_{ij}=x^{old}_{ij}+u(x^{old}-x_{kj})
\end{split}
\end{equation}
The probability of onlooker bees for choosing a food source:
\begin{equation} \label{MyEq15b}
\begin{split}
P_i=\tfrac{fitness_1}{\sum^{E_b}_{i=1}fitness_i}
\end{split}
\end{equation}
where $i=1,2,$....,$E_b$ is the half of the colony size, $j=1,2,$....,$D$ and $j$ is the number of positions with D dimension where $D$ refers to the number of parameters to be defined, $fitness_i$  is the fitness function, $k$ is a random number where $k\in (1,2,....E_b)$, $u$ is a random number between 0 and 1. \\*[.3pc]
\indent
ABC is mapped into the proposed deployment algorithm by updating the $L_i$ obtained in the first phase of the deployment. Owing to the fact that the second least eigenvalue ($\lambda_2$) of $L$ gives the algebraic connectivity of the network, the connectivity level of the backbone can be preserved within a desired limit throughout the optimization procedures. It must be noted that $L$ is symmetric and its diagonal elements are equivalent to the summation of the absolute value of row or column elements. Using the probabilistic communication model, the connectivity can be intuitively given as the probability that two devices separated by a distance $d$ can communicate with each other, and it is given by:
\begin{equation} \label{MyEq15c}
\begin{split}
P_c = K e^{-\mu d^\gamma}
\end{split}
\end{equation}

where $d$ is the Euclidian distance between the transmitter and receiver, $\gamma$ is the path loss exponent calculated based on experimental data, $\mu$ is a random variable that follows a log-normal distribution function with zero mean and variance $\sigma^2$ to describe signal attenuation effects in the monitored site, and $K$ is a constant calculated based on the transmitter, receiver and monitored site mean heights. Therefore, the probabilistic connectivity $P_c$ depends not only on the distance separating the wireless nodes but also on the surrounding obstacles and terrain, which can cause shadowing and multipath effects, $\mu$. See \cite{Turjman:2012} for more detailed explanation. In this work, the network connectivity within the bound $ \lambda_2^{min} \leq \lambda_2 \leq \lambda_2^{max}$, and in the simulation is constrained where $\lambda_2^{min} = 0.4$ and $\lambda_2^{max} = 0.6$ are set as the lower and upper bounds respectively. However, these parametric bounds can be tuned to a more desired threshold. For example in environmental  monitoring application, the connectivity bounds can be selected as $ 0.1 \leq \lambda_2 \leq 0.3$ as shown in \cite{Turjman:2012}, which focuses mainly on optimizing the connectivity level.
\\\\*[-.6pc] 
\indent
In the second phase, relay nodes ($SPRN$) are deployed and the proposed algorithm updates the matrix $L$ as shown in Eq. (\ref{MyEq11}) and determines the positions of these SPRN among candidate positions on the Grid structure where the objective function is minimal. The objective function is chosen as given in Eq. (\ref{MyEq16}).
\begin{equation} \label{MyEq16}
\begin{split}
fit = n \sum ^n_{i=2}(\tfrac{1}{\lambda_i})
\end{split}
\end{equation}
The objective function of the optimization is a minimization function subject to constraints on cost and $\lambda_2$. In this study, ABC algorithm is developed as a tool to optimize the value of the elements in the upper triangle of the matrix and simultaneously updating the diagonal elements. Each element in the upper triangle of the matrix excluding the diagonal will be optimized to be either 0 or -1 depending on whether a link is established or terminated. The connectivity constraints must be satisfied before accepting the selected positions. The average distance between the RN/CH is evaluated from the updated Laplacian matrix in conformity with  Eq. (\ref{MyEq7}). The total number of rounds is then evaluated using Eq. (\ref{MyEq12}) through Eq.(\ref{MyEq15}).
\\\\*[.4pc]
\indent
One of the critical variable in the energy consumption model is the distance between the transmitter and the receiver as shown in Eq.(\ref{MyEq14}). Owing to direct proportion, minimizing $W(G)$ will consequently minimize the average inter-node distances $(\mu_w(G))$, which inevitably extends the lifetime of the network. This makes it obvious why Wiener index ($W(G)$)is formulated as the objective function of the optimization problem. The dimension of $L$ and the number of its eigenvalues are determined by the total number of relay nodes, cluster heads and base station.\\*[.4pc]
\indent
The Laplacian matrix is updated by placing an incremental number of the SPRNs on the candidate positions. Selecting the best positions that yield the minimum average inter-node distance is the main challenge. The proposed algorithm executes a random search among the candidate positions for any specified number of SPRNs. Then the fielder value of $L$ is calculated and checked whether it falls within the pre-defined range. Else, the algorithm discards the solution.\\*[.4pc]
\indent
At the end of the whole optimization process, the average distance is evaluated from the best set of eigenvalues. The optimal deployment solution with the least average distance will then be apparent in the updated $L$. Connected nodes in the neighborhood of these SPRNs will then be visible by observing the optimal $L$.
\begin{algorithm}
	\caption{Deployment of SPRN}
	\label{alg:eldwcc}
	\begin{algorithmic}[1]
		\Function{ILDCC}  {B: Constructed Backbone using CHs, FPRNs \& BS}
		\State {\bf Input}:
		\State {A set B of the CHs, FPRNs and BS nodes positions on grid structure.};
		\State {\bf Output}:
		\State A set $X_P$ of the SPRNs coordinates maximizing lifetime of B with practical connectivity and cost constraints;
		\State {\bf begin}
		\State $L_i$ $=$ Initial Laplacian matrix of $B$;
		\State $I_R$ $=$ number of rounds B can stay operational for
		\State $A_i$ $=$ adjacency matrix matching vertex i on the grid
		\State Evaluate $\mu_w(G)$;
		\State $E_R$ $=$ extra rounds achieved by allocating RN at vertex i;
		\State $WC$ $=$ Solution of ABC in Figure \ref{abc}
		\EndFunction
	\end{algorithmic}
\end{algorithm}

\section{Simulation Results} \label{sec5}
In the first simulation, 9 CHs and 1 Base Station are deployed on known grid locations at a very short distances to their cluster members. The effectiveness of the proposed algorithm is validated using MATLAB. The number of deployed relay nodes is varied between 10 and 50. For easier comparison, the following metrics are used:
\begin{enumerate}
	\item Number of deployed relay nodes, which indicates the cost effectiveness of the strategies and
	\item Number of rounds which gives the duration the relay nodes can stay connected before partitioning. This also signifies the lifetime (in rounds) a wireless sensor network can stay operational after the deployment of RNs.\\*[-.2pc]
\end{enumerate}
The value of normalized $\Delta\mu(G)$ used in the simulation is 0.1. In order to benchmark the proposed strategy with some existing solutions, the SP3D algorithm has been implemented.
\\*[.6pc]
\indent
It must be noted that the lifetime derivations in Eq. (\ref{MyEq12})-(\ref{MyEq15}) in the foregoing section give the absolute number of rounds the network can stay operational. However, the number hugely depends on many factors such as  the network area, the specified density of active sensors, the data generation rate and the amount of data relayed on behalf of other  nodes. For the purpose of giving a global picture of the energy efficiency associated with different deployment schemes, the normalized average internode distance, lifetime and connectivity are proposed. \\*[.4pc]
The number of generations used in this experiment is 200 and the population size varies with the network size. The population size increases with increased network size. The experiment was repeated eight times with different initial populations for each network size. However, different but closely related observations were recorded. While running the experiments with a network size equals 20, we observed that the objective function did not change significantly as the number of iterations increases to the neighborhood of 200. It must be remarked that the network size refers to the total number of deployed RNs/CHs, both in the first and second phases plus the base station. The parameters to be optimized are the off-diagonal elements (upper and lower diagonal elements) of the Laplacian  matrix constructed from the network architecture, and they also represent the number of employed bees. The population size is the number of sources that the bees will visit to find better solution or better source. Table \ref{table:result1} presents the number of optimized parameters and the number of iterations as the network size increases from 20 to 60. The results of experiment are presented in Tables \ref{table:result2}-\ref{table:result6}. It was observed in Tables \ref{table:result2}-\ref{table:result6} that the average values obtained for eight experiments is close to the value observed for each individual experiment. This is a significance that the proposed deployment strategy is robust even though the ABC algorithm is executed only for 200 iterations.\\*[.4pc]
\begin{table}[t] \label{Table1}
	\setlength{\tabcolsep}{1pt}
	\caption{Numerical value for simulations} 
	\centering 
	\small
	\begin{tabular}{|c| c| c| c|} 
		\hline\hline 
		Parameters & Value & Parameters & Value\\ [0.5ex]
		\hline\hline 
		$n_c$ & $110$ & $L$ & 512 bits \\[1ex] 
		$J_a$ & $50 \times 10^{-7} J$ & $E_i$ & 15.4 J\\[1ex]
		$\epsilon_1$ & $50 \times 10^{-9}$ J/bit & $T$ & 100(p/r) \\[1ex]
		$\epsilon_2$ & $10 \times 10^{-12} J/bit/m^2$ & $R$ & 100\\[1ex]
		$\gamma$ & $4.8$ & $A$ & 10\\[1ex]
		$\beta$ & $50\times 10^{-9} J/bits$ & $r$ & 100m\\[1ex]
		\hline\hline 
	\end{tabular}
	\label{table:nomen} 
\end{table}

\begin{table*}[!t] \label{Table1}
	\setlength{\tabcolsep}{1pt}
	\caption{ABC Optimization} 
	\centering 
	\small
	\begin{tabular}{|c| c| c| c|} 
		\hline 
		Network Size&Number of Optimized Parameters&Population Size&Number of Generations\\ [0.5ex]
		\hline\hline 
		20 & 190 & 400 & 200 \\[1ex] 
		30 & 435 & 600 & 200\\[1ex]			
		40 & 780 & 800 & 200 \\[1ex]			
		50 & 1225 & 1000 & 200\\[1ex] 			
		60 & 1770 & 1200 & 200\\[1ex]			
		\hline\hline 
	\end{tabular}
	\label{table:result1} 
\end{table*}

\begin{table*}[!t]
	\setlength{\tabcolsep}{1pt}
	\caption{Network Size ($N=20$)} 
	\centering 
	\small
	\begin{tabular}{|c| c| c| c|c| c| c| c|c|c|} 
		\hline 
		Experiment No &   1   &     2    &   3    &   4  &   5   &   6   &   7  &   8  &  Average\\ [0.5ex]
		\hline\hline 
		$W(G)$ &   20.0070    &     20.4224   &  	22.0462    &  20.0433 &  19.4899   &   20.4215   &   20.6681  &  20.0145  &  20.3891 \\[1ex] 
		\hline
		$\mu(G)$ &  10.5155   &    11.1749   &   11.2212    &  11.7035  &   11.7047   &   12.0180   &   13.7755  &  11.1844  &  11.6622	\\[1ex]   	   	   	   	   	   	   				 \hline
		$E_p$ & 0.5679   &   0.6340    &   0.6388    &   0.6903  &   0.6905   &   0.7253   &   0.9420  &  0.6350  &  0.6905 \\[1ex]	  	  	  	  							 \hline
		$T_R$ &  4.2745   &    3.5761   &   3.5297    &   3.0659  &   3.0648   &  2.7814   &   1.4180  &   3.5665 &  3.1596\\[1ex]
		\hline											
		$\lambda_2$ &  0.5978   &     0.5958   &   0.5833    &  0.5968  &  0.5985   &  0.5981   &  0.5995 &  0.5973 &  0.5959\\[1ex]											 
		\hline\hline 
	\end{tabular}
	\label{table:result2} 
\end{table*}

\begin{table*}[!t] \label{Table1}
	\setlength{\tabcolsep}{1pt}
	\caption{Network Size ($N=30$)} 
	\centering 
	\small
	\begin{tabular}{|c| c| c| c|c| c| c| c|c|c|} 
		\hline 
		Experiment No &   1   &     2    &   3    &   4  &   5   &   6   &   7  &   8  &  Average\\ [0.5ex]
		\hline\hline 
		$W(G)$ &   49.9406    &     52.5264   &  	51.9119    &  48.7153 &  50.7329   &   48.9233   &   49.6524  &  48.4960  &  50.1123 \\[1ex] 
		\hline  								
		$\mu(G)$ &  10.8104   &    12.2508   &   11.9085    &  10.1279  &   11.2518   &   10.2437   &  10.6499  &  10.0057  &  10.9061	\\[1ex]   	   	   	   	   	   	   				 \hline
		$E_p$ & 0.5969   &   0.7519    &   0.7130    &   0.5311  &   0.6420   &   0.5419   &   0.5810  &  0.5198  &  0.6097 \\[1ex]	  	  	  	  							 \hline
		$T_R$ &  8.0350   &    5.9746   &   6.4236    &   9.1859  &   7.3542   &  7.3542   &   8.9817  &   8.2946 &  8.8762\\[1ex] 																			 \hline    								
		$\lambda_2$ &  0.5746   &     0.5628   &   0.5750    &  0.5673  &  0.5680   &  0.5724   &  0.5636 &  0.5758 &  0.5699\\[1ex]											 
		\hline\hline 
	\end{tabular}
	\label{table:result3} 
\end{table*}
\begin{table*}[!t]
	\setlength{\tabcolsep}{1pt}
	\caption{Network Size ($N=40$)} 
	\centering 
	\small
	\begin{tabular}{|c| c| c| c|c| c| c| c|c|c|} 
		\hline 
		Experiment No &   1   &     2    &   3    &   4  &   5   &   6   &   7  &   8  &  Average\\ [0.5ex]
		\hline\hline 
		$W(G)$ &   71.6643	&78.7479	&71.8042	&72.6604	&73.9323	&72.1253	&74.3875	&73.1210	&73.5554 \\[1ex] 
		\hline  								
		$\mu(G)$ &  4.3395	&6.5401	&4.3830	&4.6489	&5.0440	&4.4827	&5.1855	&4.7920	&4.9270	\\[1ex]   	   	   	   	   	   	   				
		\hline
		$E_p$ & 0.1337	& 0.2547	& 0.1357	& 0.1488	& 0.1689	& 0.1406	& 0.1764	& 0.1560	& 0.1644 \\[1ex]	  	  	  	  							 \hline
		$T_R$ &  37.0847	&25.2982	&36.8020	&35.1218	&32.7765	&36.1619	&31.9786	&34.2523	&33.6845\\[1ex] 																			 \hline    								
		$\lambda_2$ &  0.5626	&0.5785	&0.5635	&0.5682	&0.5734	&0.5613	&0.5753	&0.5682	&0.5689\\[1ex]											
		\hline\hline 
	\end{tabular}
	\label{table:result4} 
\end{table*}
%
\begin{table*}[!t]
	\setlength{\tabcolsep}{1pt}
	\caption{Network Size ($N=50$)} 
	\centering 
	\small
	\begin{tabular}{|c| c| c| c|c| c| c| c|c|c|} 
		\hline 
		Experiment No &   1   &     2    &   3    &   4  &   5   &   6   &   7  &   8  &  Average\\ [0.5ex]
		\hline\hline 
		$W(G)$ &  99.1732	&103.4672	&99.7243	&105.1815	&106.1542	&100.7085	&102.2467	&102.9979	&102.4567 \\[1ex] 
		\hline  								
		$\mu(G)$ &  1.2744	&2.1238	&1.3835	&2.4629	&2.6553	&1.5781   	&1.8824   	&2.0310	&1.9239	\\[1ex]   	   	   	   	   	   	   				
		\hline
		$E_p$ & 0.0124	&0.0412	&0.0159	&0.0537	&0.0610	&0.0223   &0.0327	&0.0379	&0.0346 \\[1ex]	  	  	  	  							
		\hline
		$T_R$ &  81.8460	&70.0691	&80.2143	&65.9172	&63.6861	&77.3925	&73.2067	&71.2569	&72.9486\\[1ex] 																			 \hline    								
		$\lambda_2$ &  0.5776	&0.56381	&0.5761	&0.5653	&0.5510	&0.5649	&0.5643	&0.5762	&0.5674\\[1ex]											
		\hline\hline 
	\end{tabular}
	\label{table:result5} 
\end{table*}

\begin{figure}[!h]
	\centering
	\includegraphics[scale=0.45]{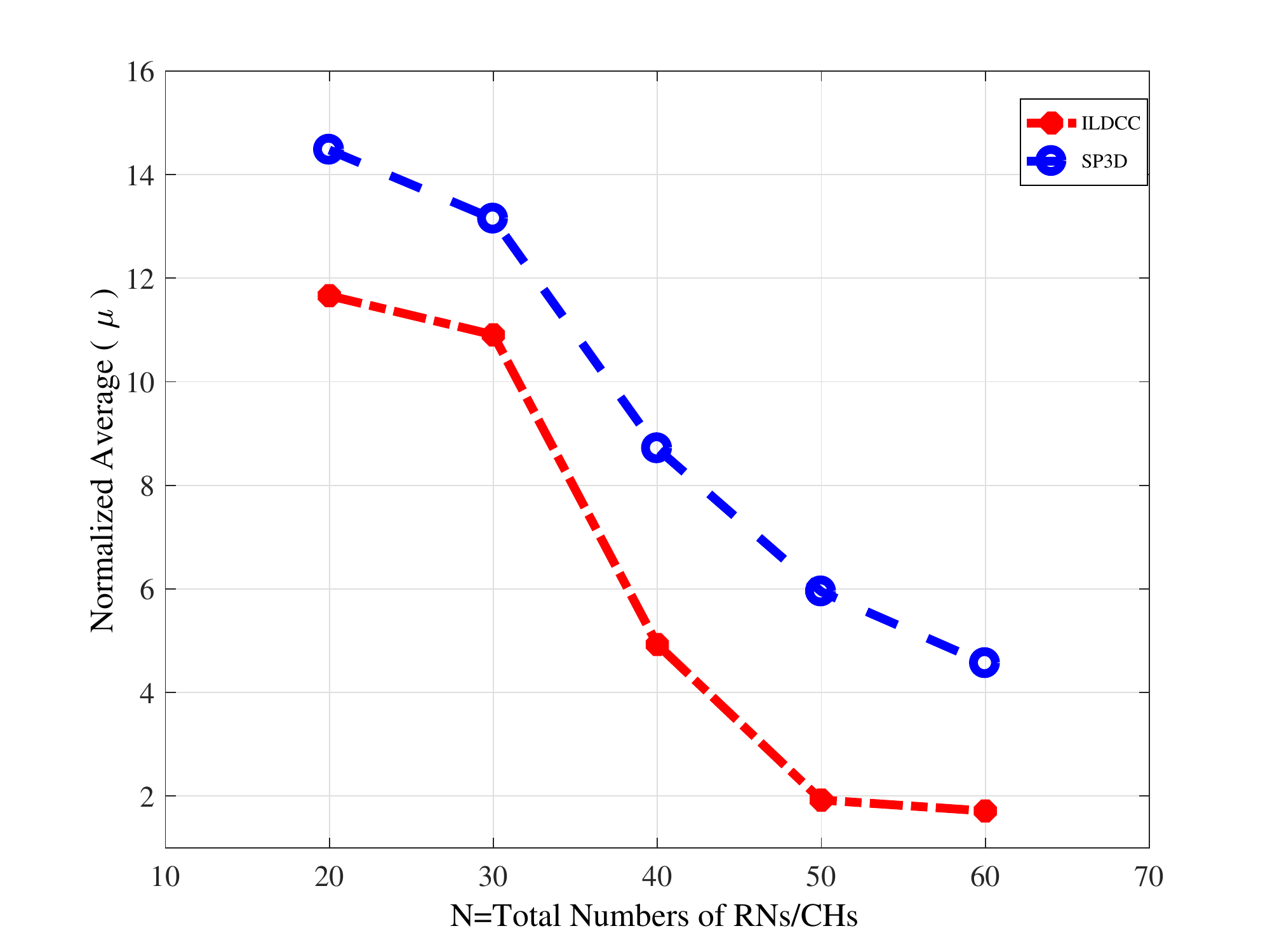}
	\caption{Normalized Average distance vs number of nodes}\label{NET_ABC_Muo}
\end{figure}

\begin{figure}[!h]
	\centering
	\includegraphics[scale=0.45]{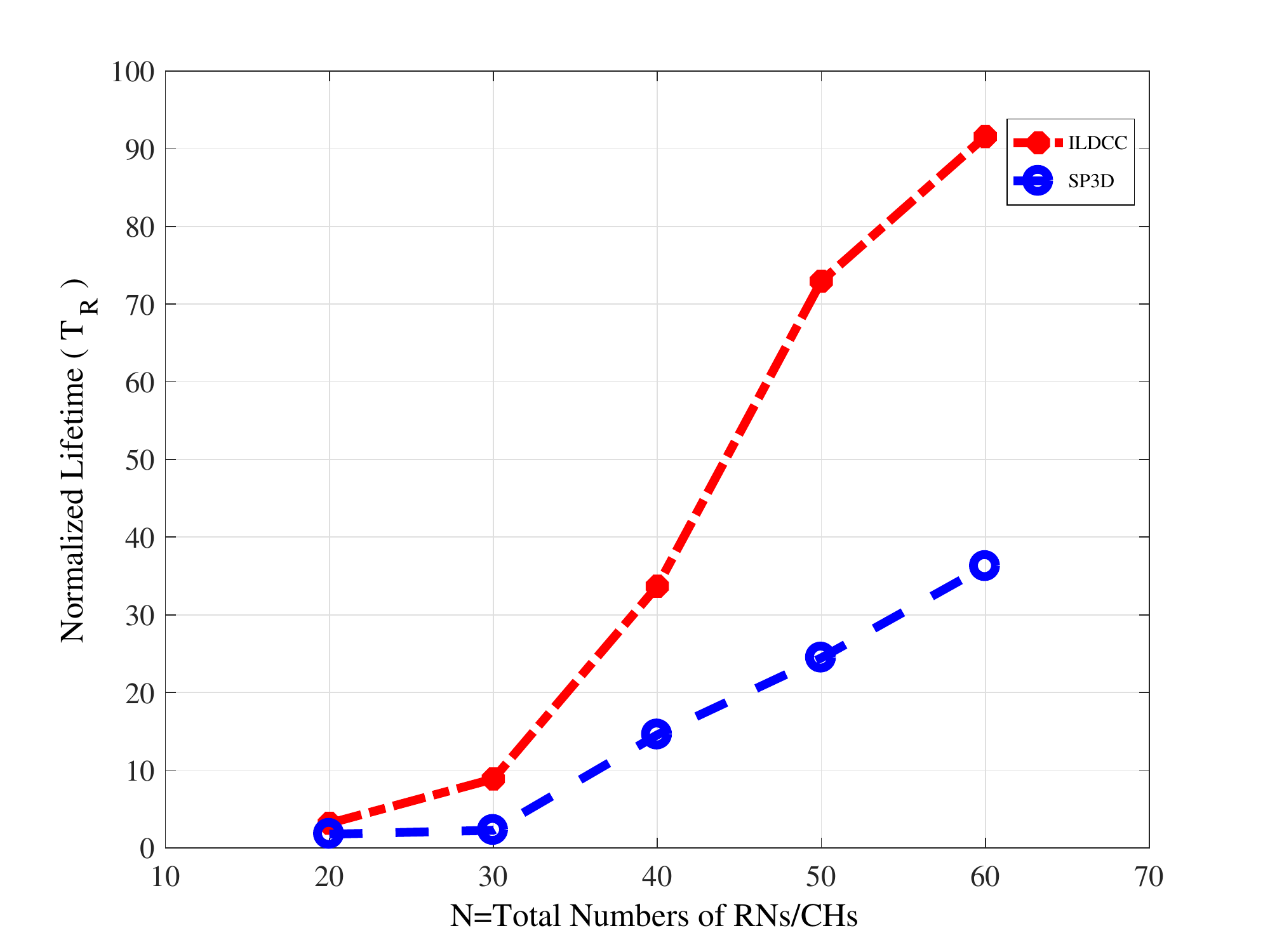}
	\caption{Lifetime vs. number of nodes using packets per round=30.}\label{NET_ABC_TR}
\end{figure}

\begin{figure}[!h]
	\centering
	\includegraphics[scale=0.45]{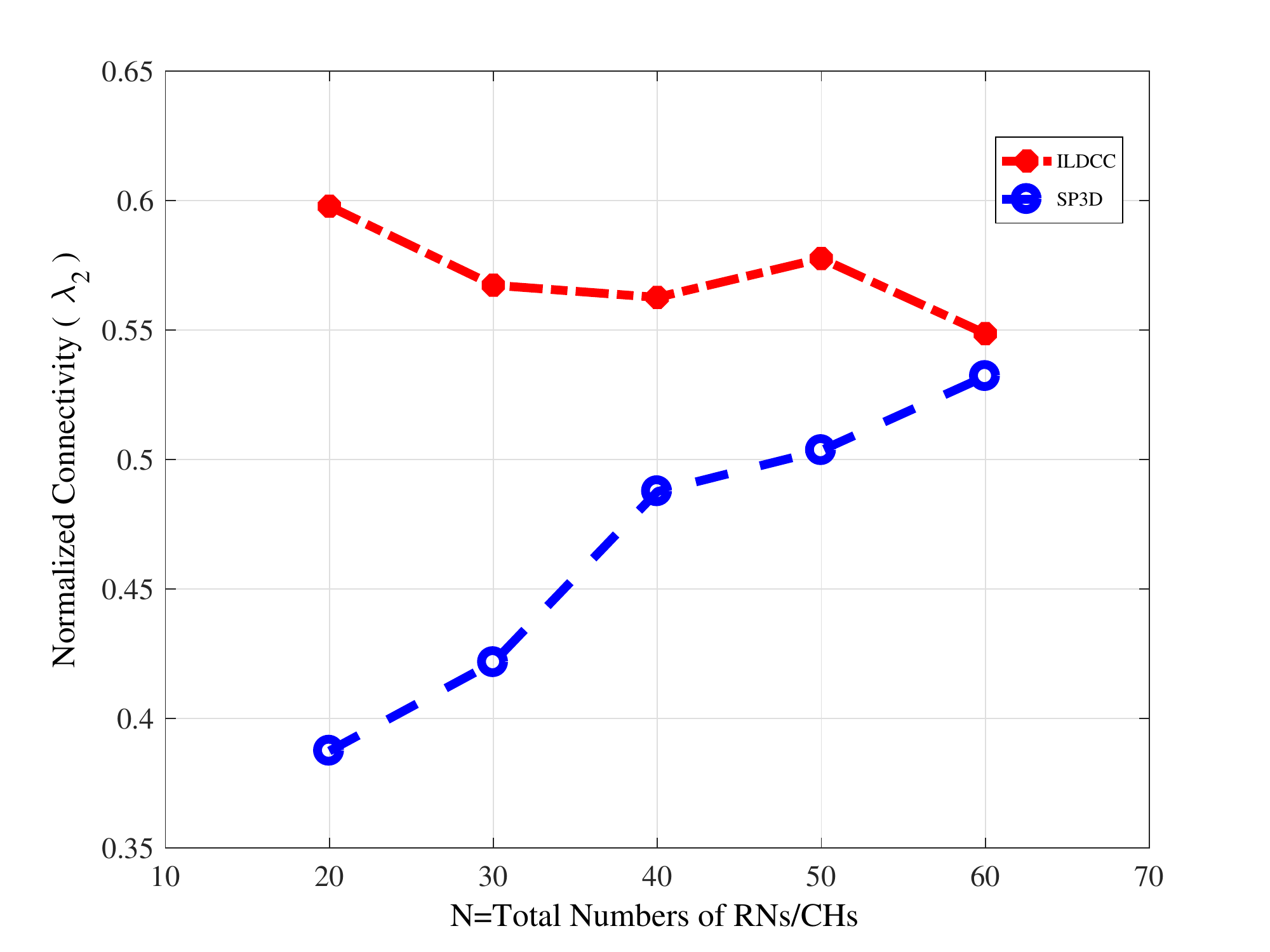}
	\caption{Connectivity vs. number of nodes }\label{NET_ABC_Lambda2}
\end{figure}

\begin{table*}[!t] \label{Table1}
	\setlength{\tabcolsep}{1pt}
	\caption{Network Size ($N=60$)} 
	\centering 
	\small
	\begin{tabular}{|c| c| c| c|c| c| c| c|c|c|} 
		\hline 
		Experiment No &   1   &     2    &   3    &   4  &   5   &   6   &   7  &   8  &  Average\\ [0.5ex]
		\hline\hline 
		$W(G)$ &  149.2980  	&146.5478  	&144.8951	&151.6236	&148.2468  	&146.9982  	&147.8522	&149.7484	&148.1513 \\[1ex] 
		\hline  								
		$\mu(G)$ &  1.8700    	& 1.4935	& 1.2672	& 2.1883	& 1.7261	& 1.5552	& 1.6721   	& 1.9316	& 1.7130	\\[1ex]   	   	   	   	   	   	   				 \hline
		$E_p$ & 0.0323   	&0.0195	&0.0122	&0.0436	&0.0273	&0.0216	&0.0255	&0.0344	&0.0270 \\[1ex]	  	  	  	  							
		\hline
		$T_R$ &  88.8886	&95.1680  	&99.1880  	&83.9494	&91.2305	&94.1052	&92.1279   	&87.9063	&91.5705\\[1ex] 																			 \hline    								
		$\lambda_2$ &  0.5542	&0.5352	&0.5486	&0.5671	&0.5724	&0.5596	&0.5371	&0.5519	&0.5533\\[1ex]											
		\hline\hline 
	\end{tabular}
	\label{table:result6} 
\end{table*}
\indent
In the first experiment, the number of backbone nodes deployed is 20 (1 BS, 9 CHs and 10 RNs) and this number is incremented by 10 up to 60, for subsequent experiments. As seen from the results in Figure~\ref{NET_ABC_Lambda2}, the overall connectivity of the network was preserved within the specified range. As expected, the average distance in the connected backbone decreases as the number of relay nodes increases in the two strategies. Since ILDCC is an enhancement of SP3D, a smaller average internode distances is achieved as depicted in Figure~(\ref{NET_ABC_Muo}). ILDCC achieves around 50\% decrement in the normalized average distance than SP3D when $N=40$. Mathematically, the number of rounds a network can stay operational depends on the absolute distance between two communicating nodes, as shown in Eq. (\ref{MyEq14}). It can be seen that the lifetime of the network is greatly improved on implementing ILDCC as depicted in Figure~\ref{NET_ABC_TR} which conforms with the law of science.\\*[-.1pc]

\begin{figure}[!h]
	\centering
	\includegraphics[scale=0.45]{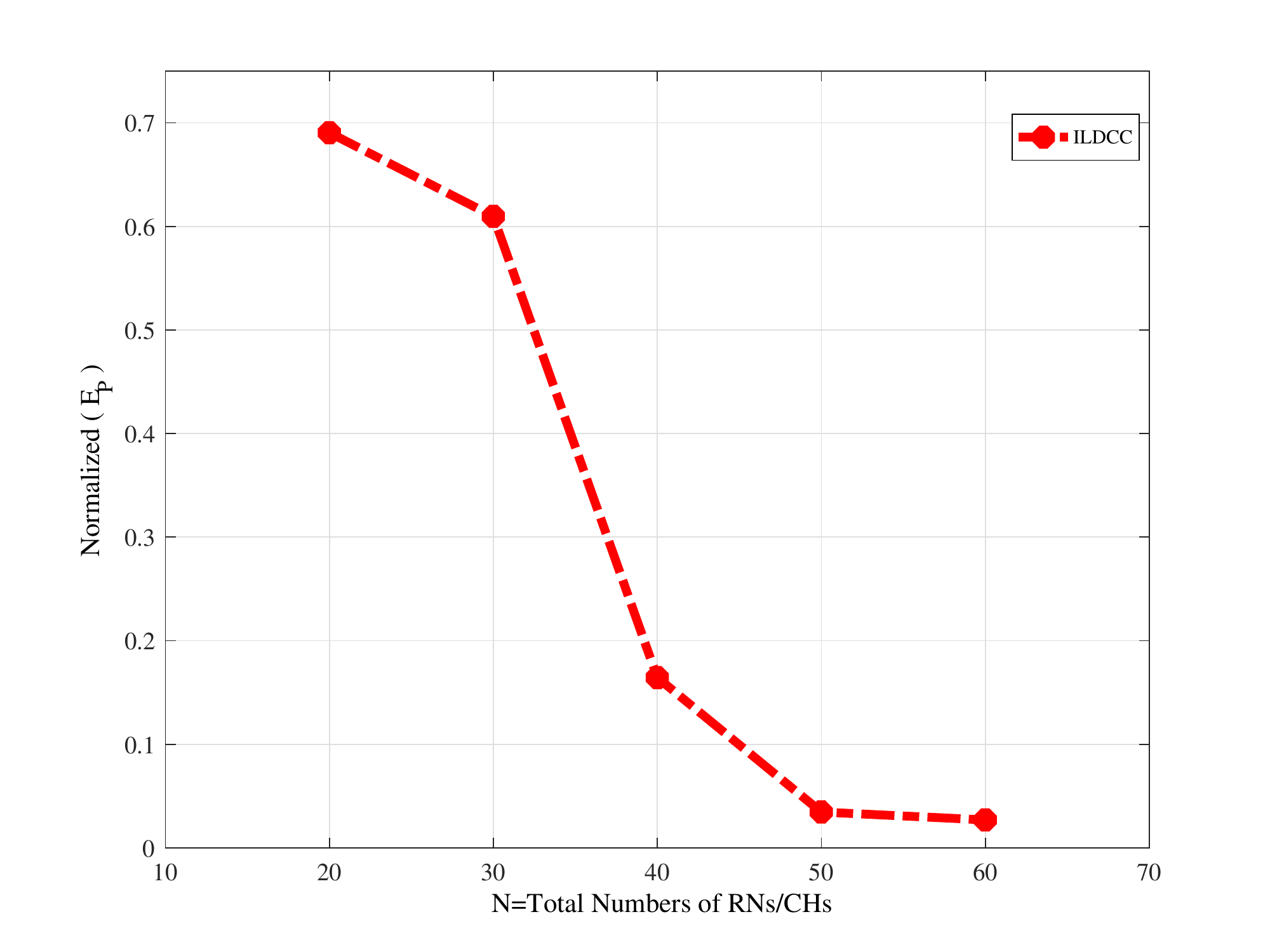}
	\caption{Lifetime vs. Network load.}\label{NET_ABC_EP}
\end{figure}
\begin{figure}[!h]
	\centering
	\includegraphics[scale=0.45]{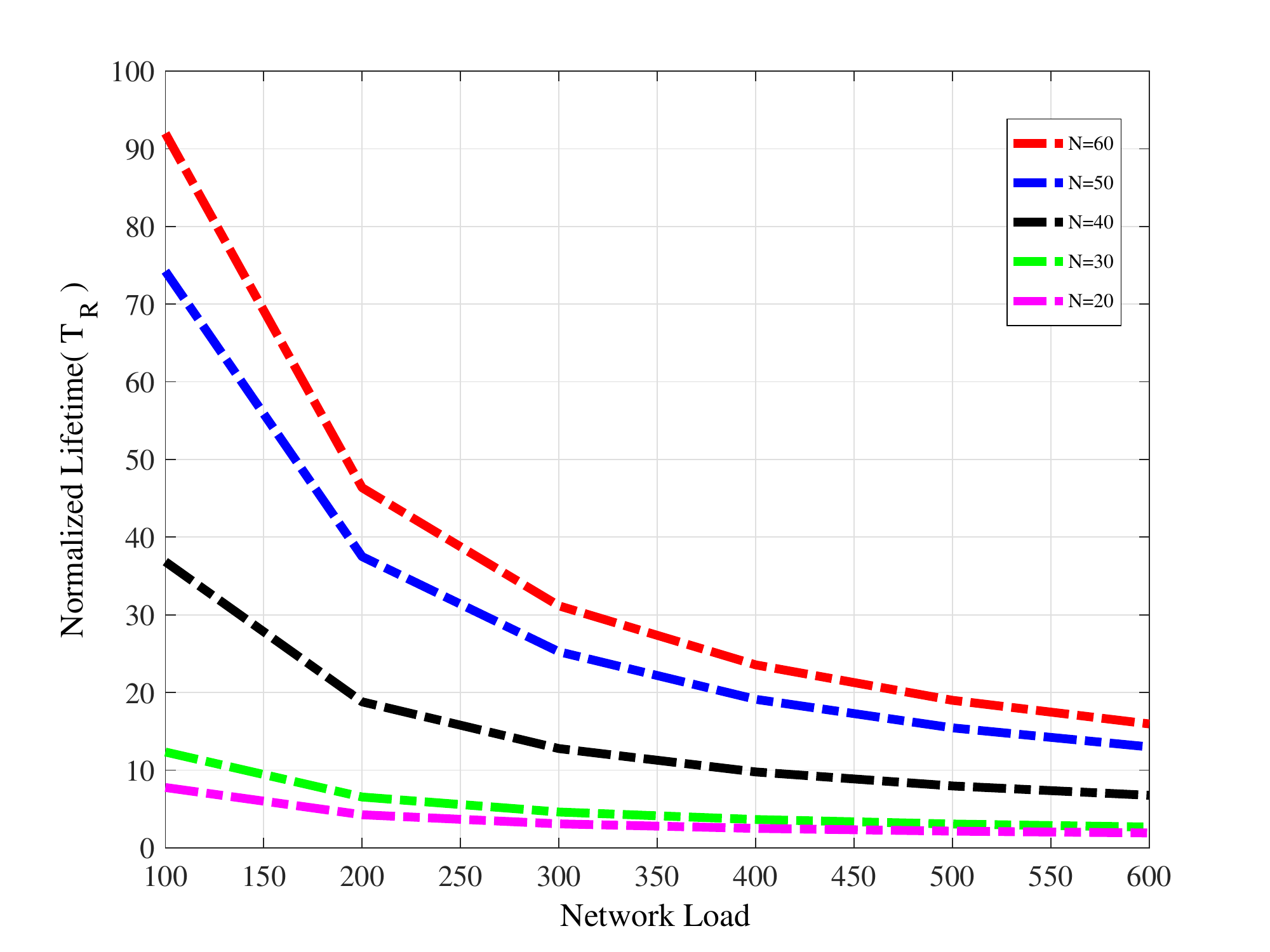}
	\caption{Lifetime vs. Network load}\label{NET_ABC_NetLoad}
\end{figure}
This phenomenon of having lifetime improved by decreasing the communicating distance is valid since the energy consumption of two communicating nodes is proportional to the square of Euclidean distance between them. Lastly, the scenario when more sensor nodes are deployed into sensing field and each of the CH/RN has more data to relay to the BS has been investigated. This accounts for more traffic generated as a result of extra data collected by the newly added sensors. In Figure~\ref{NET_ABC_NetLoad}, the traffic level was varied for different network size and it can again be observed that, the more the traffic, the larger the energy used up for transmitting and receiving packets. Apparently, increasing the energy used by the two communicating relay nodes shortens the lifetime of the overall network. To achieve a lifetime of at least 10 rounds with a traffic level of 600ppr, a minimum of 50 relay nodes have to be deployed in handling that level of traffic.
\section{Conclusion} \label{sec6}
The problem of enhancing lifetime of  wireless sensor networks is studied in this paper. Due to the fact that the energy consumption of a relay node is dependent on both the transmission radius and the traffic level, the problem is formulated as an optimization problem. A two layered, two-phase deployment architecture is proposed to avoid the NP-hard problem in 3-D settings. The backbone of the network is set up in the first phase using the minimum spanning tree protocol while Artificial Bee Colony is used to optimize the SPRN positions for lifetime maximization. Though the deployment strategy is  cost effective and easy to implement, it however has some drawbacks. For instance, increasing the RNs in a network elevates the possibility of collision and interference. With the help of data link layer protocols, the effects of collision and interference can be minimized.\\*[.4pc]
\indent
In future, the proposed technique will be benchmarked with other ABC-based algorithms in terms of the complexity and convergence. Also base station mobility will be considered to enhance the overall lifetime of the wireless sensor networks.
\section*{Acknowledgement}
The authors would like to acknowledge the support provided by King Abdulaziz City for Science and Technology (KACST) through the Science and Technology Unit at King Fahd University of Petroleum and Minerals (KFUPM) for funding this work through project No. 14-ENE265-04 as a part of the National Science, Technology and Innovation Plan (NSTIP).
\bibliographystyle{chicago}
\bibliography{sensorbibnew}

\end{document}